\documentclass[12pt,preprint]{aastex}

\slugcomment{Submitted to ApJ Letters} 
\shorttitle{Characteristic Lengths of Turbulent Magnetic Field}
\shortauthors{Cho \& Ryu}

\font\mbf = cmmib10 scaled\magstep1
       \font\mbfs = cmmib10 \font\mbfss = cmmib10 scaled 833
       \def\bmit{\fam9 }
\font\msybf = cmbsy10 scaled\magstep1
       \font\msybfs = cmbsy10 \font\msybfss = cmbsy10 scaled 833
       \def\bmsy{\fam10 }

\begin{document}

\title{Characteristic Lengths of Magnetic Field
in Magnetohydrodynamic Turbulence}
 
\author{Jungyeon Cho and Dongsu Ryu\footnote{Corresponding author.}}
\affil{Dept. of Astronomy and Space Science,
       Chungnam National University, Daejeon, Korea:
       cho@canopus.cnu.ac.kr, ryu@canopus.cnu.ac.kr}

\begin{abstract}

In the framework of turbulence dynamo, flow motions amplify a weak seed
magnetic field through the stretching of field lines.
Although the amplification process has been a topic of active research,
less attention has been paid to the length scales of magnetic field.
In this paper, we described a numerical study on characteristic
lengths of magnetic field in magnetohydrodynamic turbulence. 
We considered the case of very weak or zero mean magnetic field, which
is applicable to the turbulence in the intergalactic space.
Our findings are as follows.
(1) At saturation, the peak of magnetic field spectrum occurs at
$\sim L_0/2$, where $L_0$ is the energy injection scale, while the most
energy containing scale is $\sim L_0/5$.
The peak scale of spectrum of projected, two-dimensional field is
$\sim L_0$.
(2) During the stage of magnetic field amplification, the energy
equipartition scale shows a power-law increase of $\sim t^{1.5}$,
while the integral and curvature scales show a linear increase.
The equipartition, integral, and curvature scales saturate at $\sim L_0$,
$\sim 0.3L_0$, and $\sim 0.15L_0$, respectively.
(3) The coherence length of magnetic field defined in the Faraday
rotation measure (RM) due to the intergalactic magnetic field (IGMF)
is related to the integral scale.
We presented a formula that expresses the standard deviation of RM,
$\sigma_{RM}$, in terms of the integral scale and rms strength of the
IGMF, and estimated that $\sigma_{RM}$ would be $\sim 100$ and $\sim$
a few rad m$^{-2}$ for clusters and filaments, respectively.

\end{abstract}


\section{Introduction}

It is well established that the universe is permeated with magnetic
fields \citep[see, e.g.,][]{kron94}, and yet the origin of them is
not well understood \citep[see][and references therein]{kz08}.
The problem of cosmic magnetism can be divided into two parts -
the origin of seed fields and their amplification.
In this paper, we are concerned with the latter.

If a weak seed magnetic field is introduced into a turbulent medium,
flow motions stretch field lines and amplify the field.
Such turbulent amplification has been studied since 1950's
\citep[see, e.g.,][]{batc50,ka92,kuls97,cv00,bs05,sc07,rkcd08,cvbl09}.
The amplification goes through three stages:
(1) Eddy motions in turbulence are fastest on the smallest scale, which
is the dissipation scale.
Therefore, the stretching of magnetic field lines occurs most actively at
the dissipation scale first, and the magnetic energy grows exponentially.
(2) The exponential growth stage ends, when the magnetic energy becomes
comparable to the kinetic energy at the dissipation scale.
The follow-up stage is characterized by a linear growth of magnetic
energy, and a gradual shift of the peak of magnetic field spectrum
to larger scales.
(3) The amplification of magnetic field stops, when the total magnetic
energy becomes comparable to the kinetic energy.
A final, statistically steady, saturation stage is reached.

\citet{rkcd08} proposed a scenario in which turbulent flow motions are
induced via the cascade of the vorticity generated at cosmological
shocks during the formation of the large scale structure. 
Based on a model of turbulent amplification, they estimated the
strength of the intergalactic magnetic field (IGMF);
$\langle B\rangle \sim$ a few $\mu$G in clusters, $\sim 0.1 \mu$G
around clusters, and $\sim 10$ nG in filaments.
In their model, the intracluster medium is at a stage close to the
saturation one.
But the IGM in filaments are still at the linear growth stage.

The IGMF has been observed with Faraday rotation measure (RM); for
instance, \citet{ckb01} in clusters and \citet{xkhd06} outside clusters.
For the IGMF with $\langle{\bmit B}\rangle=0$, the mean value of RM is
expected to vanish, and the standard deviation represents the
observation \citep[see, e.g.,][]{rkb98}.
Hence,
\begin{equation}
\sigma_{RM} = 0.81~{\bar n_e}~B_{\parallel rms}~l~\sqrt{\frac{L}{l}}
~~~{\rm rad~m}^{-2}, \label{eq_1}
\end{equation}
where $n_e$, $B_{\parallel}$, and $l$ are in units of cm$^{-3}$, $\mu$G,
and pc, respectively, has been used to extract the strength of the IGMF.
Here, $B_{\parallel rms}$ is the root-mean-square (rms) strength of
line-of-sight magnetic field, $l$ is the coherence length, and $L$
is the path length.
But the formula has been applied without a clear definition of the
coherence length.

Here, we study length scales of magnetic field in magnetohydrodynamic
(MHD) turbulence with very weak or zero mean magnetic field, intended
for application to the IGMF.
Several characteristic lengths, which have been introduced in previous
works, are considered.
Characteristic lengths of initial seed fields in the early universe
could be either very large or very small, depending on how the seed
fields were generated.
However, the field structure at the linear growth and saturation stages
is independent of the seed fields as long as they are sufficiently weak
\citep[see, e.g.,][]{cvbl09}.
Therefore, the characteristic lengths should be determined by the energy
injection scale (or the outer scale) only.
And based on a scaling argument, \citet{sc07} stated that in the linear
growth stage, the scale at which the stretching is most active grows as
$l_s \propto t^{3/2}$. 

In this paper, through numerical simulations, we investigate the growth
and saturation of characteristic lengths.
Then, we show how the coherence length in the RM due to the IGMF is
defined in terms of the characteristic lengths, and present a formula
for $\sigma_{RM}$.
And based on the model of \citet{rkcd08}, we estimate $\sigma_{RM}$ for
clusters and filaments.

\section{Simulations}

We solved the incompressible MHD equations in a periodic box of size
$2\pi$ using a pseudo-spectral code.
We drove turbulence in Fourier space.
The forcing function has the following form:
${\bmit f}({\bmit x}, t)=\sum_{j=1}^{22}{\bmit f}({\bmit k}_j)
\exp(i{\bmit k}_j\cdot {\bmit x})+\mbox{complex conjugate}$, 
where ${\bmit f}({\bmit k}_j)$ is a complex vector that is perpendicular
to the wavevector ${\bmit k}_j$.
The 22 forcing components are nearly isotropically distributed in
the range $2\leq k \leq \sqrt{12}$, where $k=|{\bmit k}|$.
The phase of each forcing component randomly fluctuates, but it has a
correlation time of approximately unity.
The amplitude of each forcing component also randomly fluctuates.
On average, each forcing component injects similar amount of energy
\footnote{
The wavenumbers used are 2 (3 components), $\sqrt{6}$ (12 components),
3 (3 components), and $\sqrt{12}$ (4 components).
Therefore the peak of energy injection occurs at $k_0 \approx 2.5$,
and the energy injection scale is $L_0 \sim 2.5$.
A similar discrete sampling of forcing can be found in \citet{tcv93},
where they used 13 components.}.
In physical space this forcing ${\bmit f}({\bmit x}, t)$ corresponds to
statistically homogeneous driving on large scales.
Only the solenoidal component of the velocity field was driven\footnote{
Since we deal with incompressible turbulence, we applied solenoidal forcing.
However, in general cases, the properties of turbulence may  depend on
the nature of forcing
\citep[see, e.g.,][for compressible turbulence]{fks08}.}.
The strength of energy injection was tuned, so without magnetic field
the rms velocity becomes unity, $V_{rms} \sim 1$, at saturation.
In this representation, ${\bmit V}$ can be viewed as the velocity 
measured in units of the rms velocity of the system.
The density is unity and the magnetic field is multiplied by $\sqrt{4\pi}$
in our simulations, so ${\bmit B}$ can be viewed as the Alfv\a'en speed
in the same units.
The magnetic field consists of the uniform background field and the 
fluctuating field: ${\bmit B}= {\bmit B}_0 + {\bmit b}$.
At $t=0$, the magnetic field had either weak uniform component
(when $B_0 \ne 0$) or only random components (when $B_0 =0$;
Run $256H8$-$B_00$, see below), and the velocity had a support between
$2 \le k \le 4$ in the wavevector space.
We considered only the case where the kinetic viscosity, $\nu$, is equal
to the magnetic diffusivity, $\eta$.
See \citet{cv00} and \citet{cvbl09} for further details of simulations.

Simulations are denoted with $XY$-$B_0Z$, where $X=256$ or $512$ refers
to the number of grid points in each spatial direction, $Y=H8$ or $P$
refers to hyper (and their order) or physical viscosity/diffusivity,
and $Z$ refers to the strength of mean magnetic field.
In this paper, Runs $256H8$-$B_00$, $256H8$-$B_010^{-3}$,
$512P$-$B_010^{-3}$, and $512H8$-$B_010^{-4}$ are discussed.
In $256H8$-$B_00$, the mean field was zero, and initially the spectrum
of magnetic field peaked at $k\sim 70$.
All the runs have either very weak or zero mean magnetic field, since
we intend to apply the results to turbulence in the IGM\footnote{
The origin of seed magnetic fields in the IGM can be either cosmological
or astrophysical.
Theories in favor of cosmological origin suggest that weak seed magnetic
fields were created in the early universe or during the structure formation
era. 
Our initial magnetic field intends to mimic the cosmological origin of
seed fields.},
as noted in Introduction.
Besides the initial magnetic field configuration, Runs have either
different numerical resolution or different viscosity/diffusivity
to explore their effects.
In $512P$-$B_010^{-3}$ physical viscosity/diffusivity was used, while
in others hyperviscosity/diffusion was used to extend the inertial
range\footnote{
It is well known that hyper-dissipation causes the bottleneck effect,
unphysical flattening of energy spectrum near the dissipation range.
However, since the bottleneck effect is negligible at small wavenumbers,
we believe the effect does not alter the shape of the magnetic energy
spectrum at small wavenumbers.
Most length scales discussed in this paper rely on the shape of magnetic
energy spectrum at small wavenumbers. 
Therefore, we believe the use of hyper-dissipation does not affect our
results much.
Indeed, our results show that the length scales do not strongly depend 
on numerical resolution or forms of dissipation.}.
Runs with $256^3$ grid points are a subset of the simulations used in
\citet{cvbl09}.
For the properties of turbulence with different initial magnetic fields,
please refer an extensive study in \citet{cvbl09}.
Figure 1 shows the time evolution of $V^2$ and $B^2$ in four
simulations.
Here, the kinetic energy and magnetic energy densities are $V^2/2$ and
$B^2/2$.
Although the four simulations have different set-ups, their time
evolution looks similar. 
We can clearly see three stages of magnetic energy evolution.
In this limit of week or zero mean field, the resulting turbulence is
globally almost isotropic.

\section{Various Length Scales of Magnetic Field}

{\bf Peak scale of spectrum of magnetic field}, $L_{E(k)}$:
Figure 2a shows the spectra of velocity and magnetic field at an epoch
of saturation.
In $E_v(k)$, the peak occurs at the energy injection scale.
Note that the peak appears at $k=2$ rather that at 2.5, which is a
numerical artifact of discrete binning of $k$.
In $E_b(k)$, the peak occurs at a smaller scale.

{\bf Largest energy containing scale of magnetic field}, $L_{kE(k)}$:
With $E(k) \propto 1/k$ representing a power spectrum that contains
equal amount of energy in each decade of $k$, the peak scale of $kE(k)$
defines the largest energy containing scale.
Figure 2a shows that for the magnetic field, the peak of $kE_b(k)$
occurs at a scale even smaller than that of $E_b(k)$.

{\bf Peak scale of spectrum of projected magnetic field}, $L_{E(k)/k}$:
When a three-dimensional (3D) scalar quantity is projected onto a
two-dimensional (2D) plane, the energy spectrum of the projected quantity
becomes $E^{2D}(k) \propto E^{3D}(k)/k$ \citep[see, e.g.,][]{cl02}.
It is a bit more complicated for vector quantities, but the qualitative
behavior should be similar.
Figure 2a shows that for the magnetic field, the peak of $E_b(k)/k$
occurs at a scale close to that of $E_v(k)$.
Note that the largest energy containing scale of projected magnetic
field corresponds to the peak scale of $kE_b^{2D}(k)$, which is the
same as the peak scale of $E_b(k)$.

The time evolution and saturation of $L_{E(k)}$, $L_{kE(k)}$, and
$L_{E(k)/k}$, normalized with the energy injection scale $L_0$, are
shown in Figure 2b.
The rugged profiles are again a consequence of discrete binning
of $k$.
Two points are noteworthy.
(1) At saturation, $L_{E(k)} \sim L_0/2$, $L_{kE(k)} \sim L_0/5$, and
$L_{E(k)/k} \sim L_0$.
(2) The growth pattern of those scales in the linear growth stage is
not clear owing to the ruggedness in the profiles, but it seems to be
between $\sim t$ and $\sim t^{1.5}$.

{\bf Energy equipartition scale}, $L_{eq}$:
The energy equipartition wavenumber, $k_{eq}$, is defined by
\begin{equation}
\int_{k_{eq}}^{k_{max}} E_v(k)~dk = \int_0^{k_{max}} E_b(k)~dk.
\end{equation}
The time evolution and saturation of $L_{eq}$ are shown in Figure 3a.
We expect that the stretching of magnetic field lines is most active
at this scale.
If then, $L_{eq}$ should represent $l_s$ (see Introduction) and
follow $\sim t^{1.5}$ in the linear growth stage \citep{sc07}.
Indeed, $L_{eq}$ grows with a power-law index consistent with the
theoretical prediction in Figure 3a.
At saturation, $L_{eq}$ becomes close to $L_0$.

{\bf Integral scale}, $L_{int}$:
The length scale of magnetic field (and velocity too) is often
characterized with the integral scale, which is defined by
\begin{equation}
L_{int} = 2\pi \frac{\int E_b(k)/k~dk }{\int E_b(k)~dk}_.
\end{equation}
It is well known that the integral scale has the same order of magnitude
as the longitudinal and transversal integral scales, $L_l$ and $L_t$,
respectively, and in incompressible isotropic turbulence (with reflection
invariance on average) they are related by $L_l = 2L_t = (3/8) L_{int}$
\citep[see][]{my75}.

{\bf Curvature scale}, $L_{curv}$:
We also consider a typical radius of curvature of field lines, $L_{curv}$.
We define it as the distance $r$ at which the average correlation drops
to $1/e$,
\begin{equation}
\frac{\langle{\bmit B}({\bmit x}) \cdot {\bmit B}({\bmit x}+{\bmit r})
\rangle_{\bmit x}}{\langle{\bmit B}({\bmit x}) \cdot {\bmit B}({\bmit x})
\rangle_{\bmit x}} =\frac{1}{e}_,
\end{equation}
where the two points at separation $\bmit r$ are located along the same
magnetic field line and the average is taken over $\bmit x$.
The factor $1/e$ is an arbitrary choice.

The time evolution and saturation of $L_{int}$ and $L_{curv}$ are shown
in Figure 3b.
We calculated $L_{curv}$ only for runs with $256^3$ grid points.
Unlike $L_{eq}$, $L_{int}$ and $L_{curv}$ seem to grow linearly in
the linear growth stage (although the reason of the linear growth is
not clear).
At saturation, $L_{int} \sim 0.3L_0$ and $L_{curv} \sim 0.15L_0$.
Hence, for instance, we model the growth and saturation of $L_{int}$ as
\begin{eqnarray}
L_{int} \sim \left\{\begin{array}{ll}
(0.3/45)~L_0 \times t/t_{eddy},  & \mbox{ if $t/t_{eddy} < 45$}\\
0.3~L_0, & \mbox{ if $t/t_{eddy} \geq 45$} \end{array} \right._,
\label{eq_5}
\end{eqnarray}
where the eddy turnover time is defined with the vorticity around
the energy injection scale at saturation as
$t_{eddy} \equiv 1/\omega_{injection}$.

\section{Faraday Rotation Measure}

With $\langle{\rm RM}\rangle=0$ for the IGMF, the standard deviation of
RM is
\begin{equation}
\sigma_{RM} = 0.81~{\bar n_e} \left< \left(\int_0^L B_{\parallel} ds
\right)^2 \right>^{1/2}~{\rm rad~m^{-2}}. 
\end{equation}
Note that in this work, the density is assumed to be constant.
Here, $\int B_{\parallel} ds$ is the projected, 2D magnetic field.

Without loss of generality, we take $x$ as the line of sight direction.
The projected field can be written as
\footnote{
We considered turbulence in a cubic box of size $2\pi$, as noted in
Section 2.
That is, we assumed a flow of period $2\pi$ in the three directions
of space \citep[see][for further details]{lesi08}.
In this case, we have a usual Fourier series expansion:
$B_x({\bmit x})\approx \sum_{k_x,k_y,k_z=-N/2}^{N/2} \tilde{B}_x({\bmit k}) 
e^{i {\bmit x}\cdot {\bmit k} }$
and 
$\tilde{B}_x({\bmit k}) = \frac{1}{(2\pi)^3}
\int_0^{2\pi}dx\int_0^{2\pi}dy\int_0^{2\pi}dz  B_x({\bmit x}) 
e^{-i{\bmit x}\cdot {\bmit k}}
\approx \frac{1}{N^3} \sum_{l,m,n=0}^{N-1} 
\int B_x({\bmit x}_{lmn}) e^{-i {\bmit x}_{lmn}\cdot {\bmit k}}$,
where $N$ is the number of grid points in each side.
Here, ${\bmit k}=(k_x,k_y,k_z)$, and ${\bmit x}_{lmn}$ denotes
the coordinate of a grid point $(l,m,n)$ in the computational grid.}
\begin{eqnarray}
\int_0^{2\pi} B_{\parallel} ds \equiv \int_0^{2\pi} B_x dx
\approx \int_0^{2\pi} dx \sum_{k_x,k_y,k_z=-N/2}^{N/2} \tilde{B}_x({\bmit k}) 
e^{i{\bmit x} \cdot {\bmit k}} \nonumber \\
= 2\pi \sum_{k_y,k_z=-N/2}^{N/2} {\tilde B}_x(0,k_y,k_z)e^{i(yk_y+zk_z)},
\end{eqnarray}
where we used $\int_0^{2\pi} dx~ e^{ixk_x} =2\pi \delta_{0,k_x}$.
Then, the square-average of the projected field becomes
\begin{eqnarray}
\left< \left(\int_0^{2\pi} B_{\parallel} ds\right)^2 \right>
 = \frac{1}{(2\pi)^2} \int_0^{2\pi} dy \int_0^{2\pi} dz
\left(\int_0^{2\pi} B_{x} dx \right)^2 \nonumber \\
=  (2\pi)^2 \sum_{k_y,k_z=-N/2}^{N/2} \left| {\tilde B}_x(k_x=0,k_y,k_z)
\right|^2 \nonumber \\
\approx (2\pi)^2 \int dk_y dk_z \left|{\tilde B}_x(k_x=0,k_y,k_z)\right|^2
 =  \frac{(2\pi)^2}{2}\int \frac{E_b(k)}{k} dk.
\end{eqnarray}
In the last step, $\langle|{\tilde B}_x|^2\rangle_{k_x=0} = (1/2)
\langle|{\tilde B}|^2\rangle_{k_x=0}$ statistically in the $k_x=0$ plane
and $E_b(k) = 4\pi k^2 (|{\tilde B}|^2/2)$ were used, and
$dk_y dk_z$ was substituted with $2\pi k dk$.
Finally, using $\langle B^2\rangle/2 = \int E_b(k) dk$,
\begin{eqnarray}
\left< \left(\int_0^{2\pi} B_{\parallel} ds\right)^2 \right>
& = & \frac{\left<B^2\right> (2\pi)^2}{4}~
\frac{\int E_b(k)/k~dk }{\int E_b(k)~dk} \nonumber \\
& = & \frac{\left<B^2\right> L_{int}(2\pi)}{4}_.    \label{eq_9}
\end{eqnarray}
So far, we have assumed that the box size is $2\pi$.
When the box size is $L$ (or, when the path length is $L$)\footnote{
The periodic simulation box size can be used as a proxy for the
physical path length of the system
\citep[see, e.g., discussion in][]{lesi08}.}, 
$2\pi$'s in Equation (\ref{eq_9}) should be replaced by $L$. 
Hence, the standard deviation of RM becomes
\begin{equation}
\sigma_{RM} = 0.81~{\bar n_e}
\frac{B_{rms}\sqrt{L_{int}L}}{2}~{\rm rad~m^{-2}}.    \label{eq_10}
\end{equation}
Since $B_{\parallel rms} = B_{rms}/\sqrt{3}$, the coherence length used
in Equation (\ref{eq_1}) should be given as $l = (3/4)L_{int}$.

Equation (\ref{eq_10}) can be applied to estimate RMs due to the IGMF.
Here, we employ the model of \citet{rkcd08} for turbulence and magnetic
field in the IGM.
In clusters, turbulence is near the saturation stage with
$t/t_{eddy} \sim 30$, where $t \equiv t_{age}$, the age of the universe,
and $t_{eddy} \equiv \omega_{rms}^{-1}$.
From Equation (\ref{eq_5}), then, $L_{int}/L_0 \sim 0.2$.
If we take the energy injection scale $L_0 \sim 100$ kpc, which
is approximately the scale height of cluster core, $L_{int} \sim 20$ kpc.
With ${\bar n_e} \sim 10^{-3}$ cm$^{-3}$, $B_{rms} \sim$ a few $\mu$G,
and the path length $L \sim 1$ Mpc, we get $\sigma_{RM} \sim 100$
rad m$^{-2}$ for clusters, which agrees with the observed RMs in clusters
\citep{ckb01}.
In filaments, on the other hand, with $t/t_{eddy} \sim 10$, turbulence
is expected to be still in the linear growth stage, and
$L_{int}/L_0 \sim 1/15$.
We may take the energy injection scale $L_0 \sim 5$ Mpc, which is the
typical thickness of filaments.
It is also the typical radius of curvature of cosmological shocks in
filaments \citep{rkhj03}, which would be the major sources to drive
turbulence there.
The power spectrum of ${\bmit v}_{curl}$, the curl component of flow
motions which satisfies the relation
${\bmsy\nabla}\times{\bmit v}_{curl} \equiv {\bmsy\nabla}\times{\bmit v}$,
in the large scale structure of the universe peaks around 5 Mpc too
\citep{rk08}.
Then, $L_{int} \sim 300$ kpc.
The magnetic field strength in filaments quoted in \citet{rkcd08} is
$\langle B\rangle \sim 10$ nG.
But we note that the value depends on how it is averaged.
With the data of \citet{rkcd08},
$\langle B^2\rangle^{1/2} \sim$ a few $\times~10$ nG,
$\langle \rho B\rangle/\langle \rho\rangle \sim 0.1~\mu$G,
and $\langle (\rho B)^2\rangle^{1/2}/\langle \rho^2\rangle^{1/2} \sim$
a few $\times~0.1~\mu$G, in the warm-hot intergalactic medium
(WHIM) with $T = 10^5 - 10^7$ K which mostly composes filaments.
The average value of $\langle (\rho B)^2\rangle^{1/2}/\langle
\rho^2\rangle^{1/2}$ should be most relevant to RM.
Then, $\sigma_{RM}$ for filaments, if they intersect the line of sight
with right angle, would be
\begin{equation}
\sigma_{RM} \sim 1.5\  \left(\frac{\bar n_e}{10^{-5}{\rm cm}^{-3}}\right)
\left(\frac{B_{rms}}{0.3~\mu{\rm G}}\right) \left(\frac{L_{int}}{300 {\rm kpc}}
~\frac{L}{5 {\rm Mpc}}\right)^{1/2} {\rm rad~m^{-2}}.
\end{equation}
Normally filaments would not intersect the line of sight with right angle.
Smaller angles result in larger path lengths and so larger $\sigma_{RM}$,
and then the typical value of $\sigma_{RM}$ for filaments could be a few
rad m$^{-2}$.
We note that the values of $|$RM$|$ toward the Hercules and Perseus-Pisces
superclusters reported in \citet{xkhd06} is an order of magnitude larger
than the above value, and \citet{xkhd06} quoted the path length, $L$,
which is about two orders of magnitude larger.

\section{Summary}

Our findings are summarized: \hfill\break
(1) We studied different characteristic scales of magnetic field in
MHD turbulence with very weak or zero mean magnetic field.
They saturate at $\sim 0.1 - 1 L_0$, where $L_0$ is the energy
injection scale. \hfill\break
(2) During the linear growth stage of magnetic energy, the energy
equipartition scale follows $L_{eq} \propto t^{3/2}$, while
the integral scale follows $L_{int} \propto t$. \hfill\break
(3) The integral scale (actually $(3/4)L_{int}$) is the relevant scale
for RM.
We obtained a new formula for  the standard deviation of RM,
$\sigma_{RM}$ (see Eq.~[\ref{eq_10}]). \hfill\break
(4) We estimated $\sigma_{RM}$ for filaments as well as for clusters
of galaxies.

Finally, we note that our findings are based on a small number of
incompressible numerical simulations.
They appear to be rather insensitive to numerical resolution and forms of
dissipation.
However, when different initial conditions or different types of forcing
are used, some of the results may be different, a question which should be
investigated in the future.

\acknowledgements

We thank the anonymous referee for useful suggestions.
JC was supported in part by Korea Research Foundation (KRF-2006-331-C00136). 
DR was supported in part by Korea Science and Engineering Foundation
(R01-2007-000-20196-0).
JC and DR were also supported by Korea Foundation for International
Cooperation of Science and Technology (K20702020016-07E0200-01610).

\clearpage

\begin{figure}
\vskip -1cm
\hskip 4cm
\includegraphics[scale=0.5]{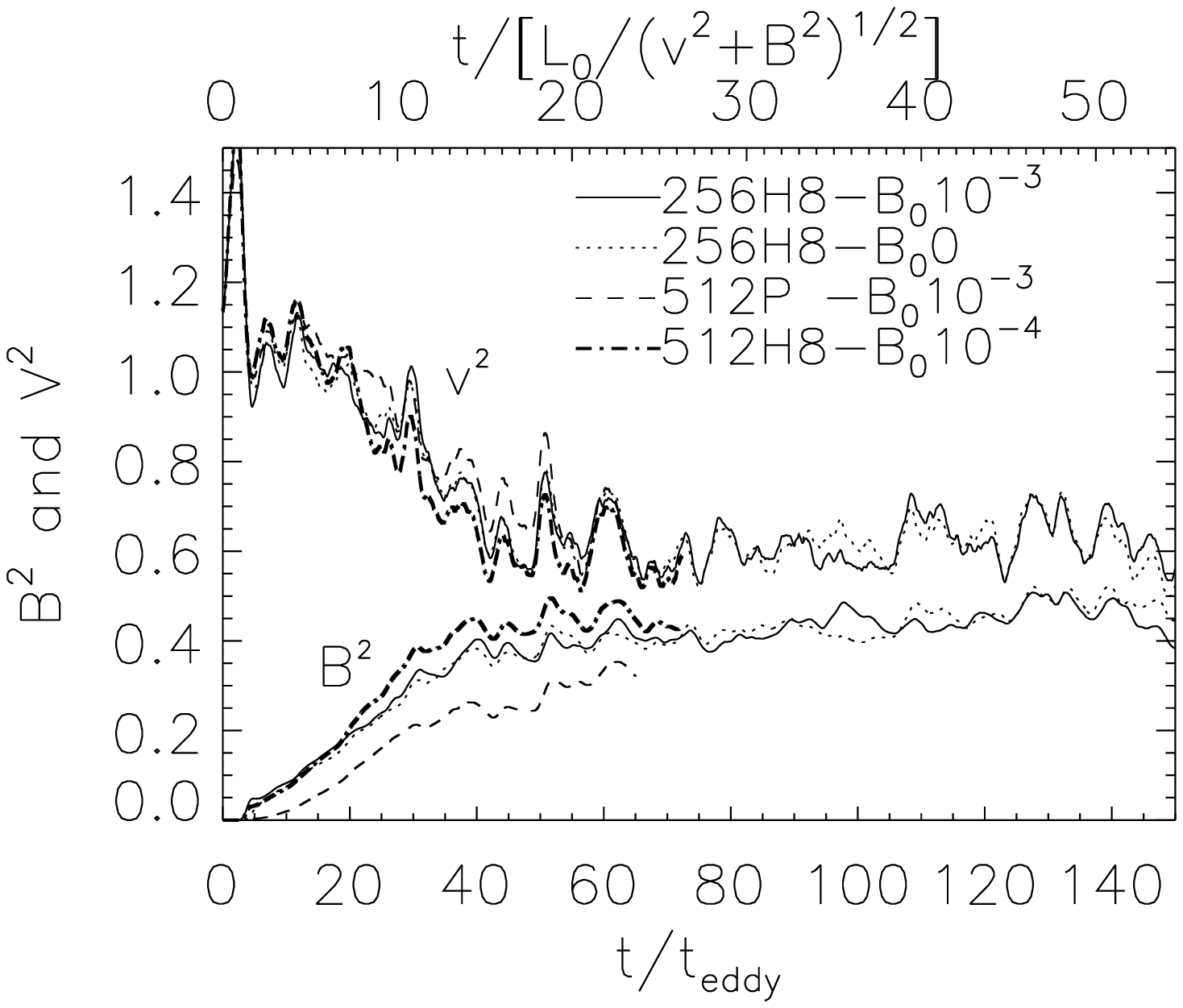}
\vskip 0cm
\caption{Time evolution of $V^2$ and $B^2$.
Time is given both in units of the eddy turnover time defined with
the vorticity around the energy injection scale at saturation,
$t_{eddy} \equiv 1/\omega_{injection}$, (bottom) as in \citet{rkcd08}
and in units of the energy injection scale divided by $\sqrt{V^2+B^2}$
at saturation (top) as in \citet{cvbl09}.}
\end{figure}

\clearpage

\begin{figure}
\vskip -1cm
\hskip 0.3cm
\includegraphics[scale=0.5]{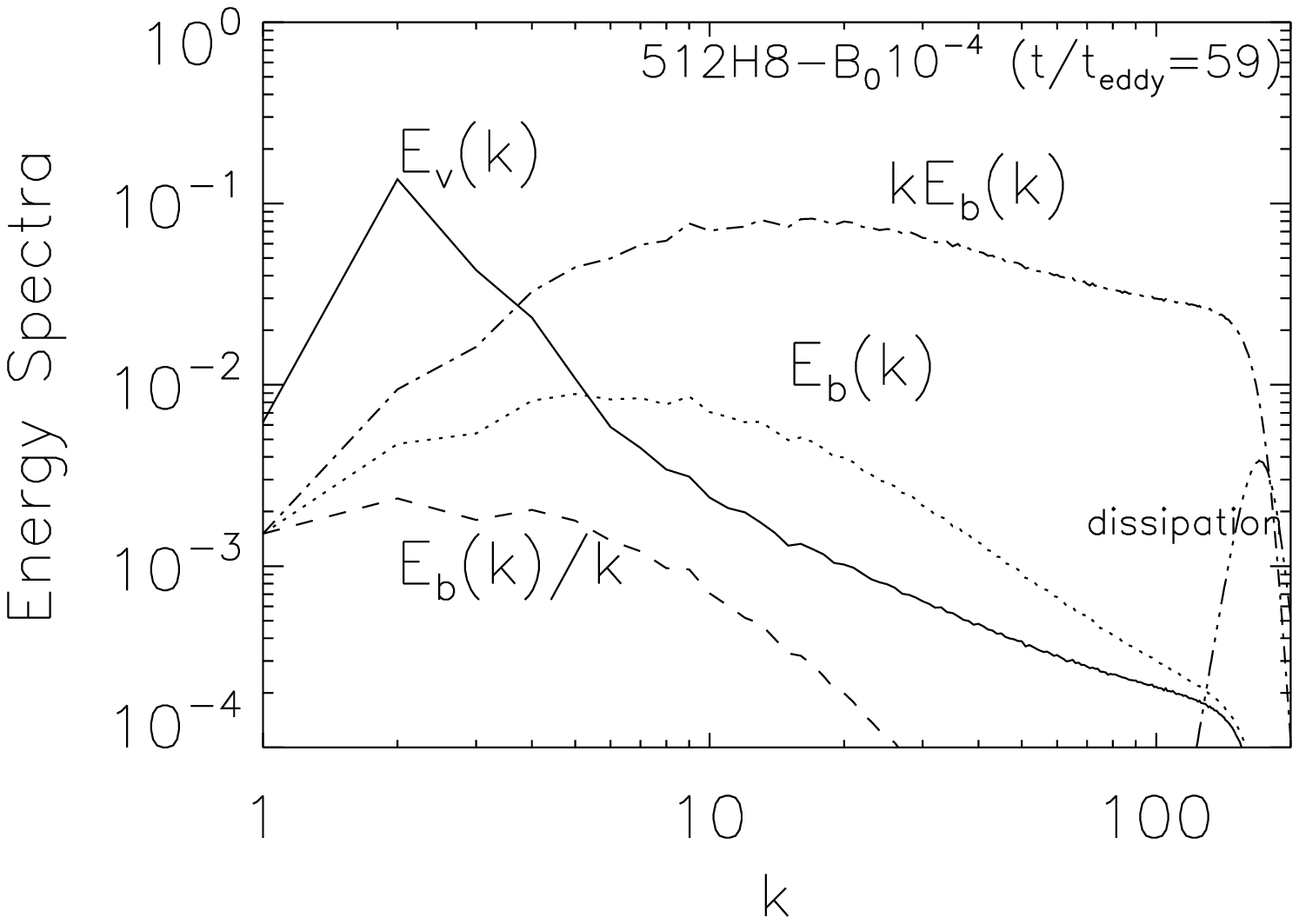}
\includegraphics[scale=0.5]{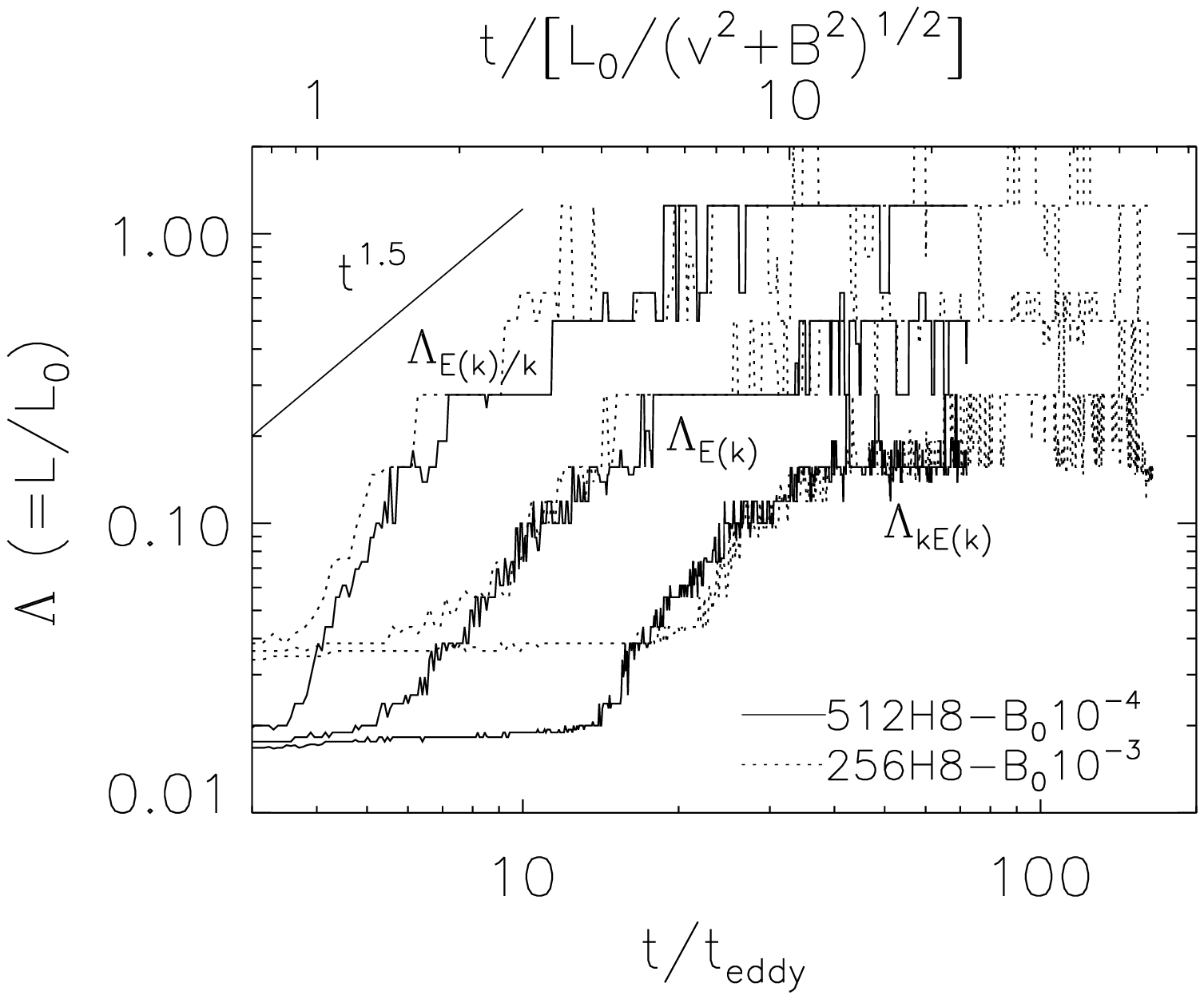}
\vskip 0cm
\caption{Left panel: Spectra of velocity, $E_v(k)$, and magnetic field,
$E_b(k)$, at $t/t_{eddy}=59$.
The dissipation spectrum is also shown.
Although hyper-dissipation affects the spectra near the dissipation scale, 
it does not strongly influence the shapes of the spectra at small wavenumbers.
Right panel: Time evolution of peak scales of $E_b(k)$, $kE_b(k)$,
and $E_b(k)/k$.
$\Lambda$'s are the scales normalized with $L_0$.}
\end{figure}

\clearpage

\begin{figure}
\vskip -1cm
\hskip 0.3cm
\includegraphics[scale=0.5]{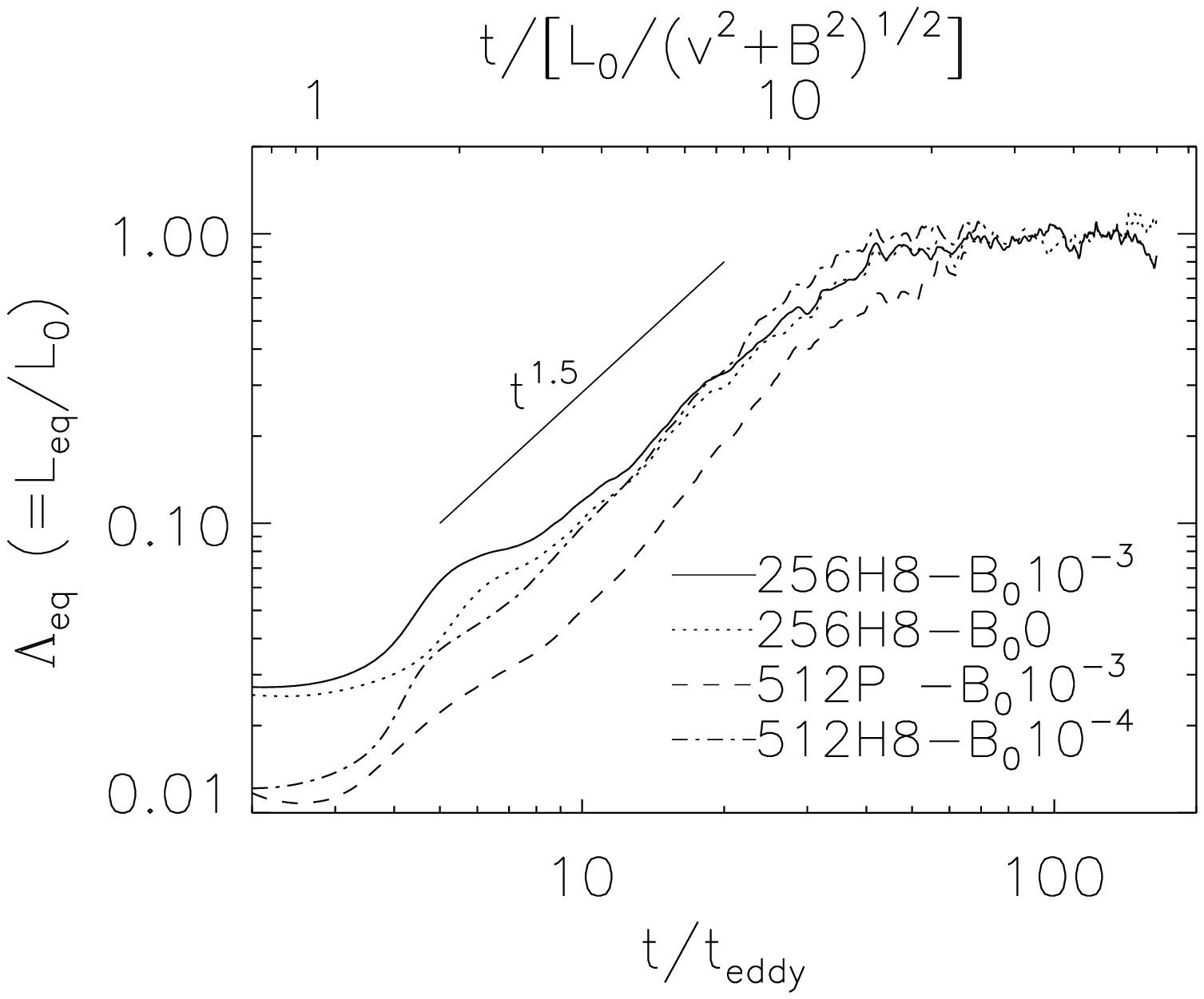}
\includegraphics[scale=0.5]{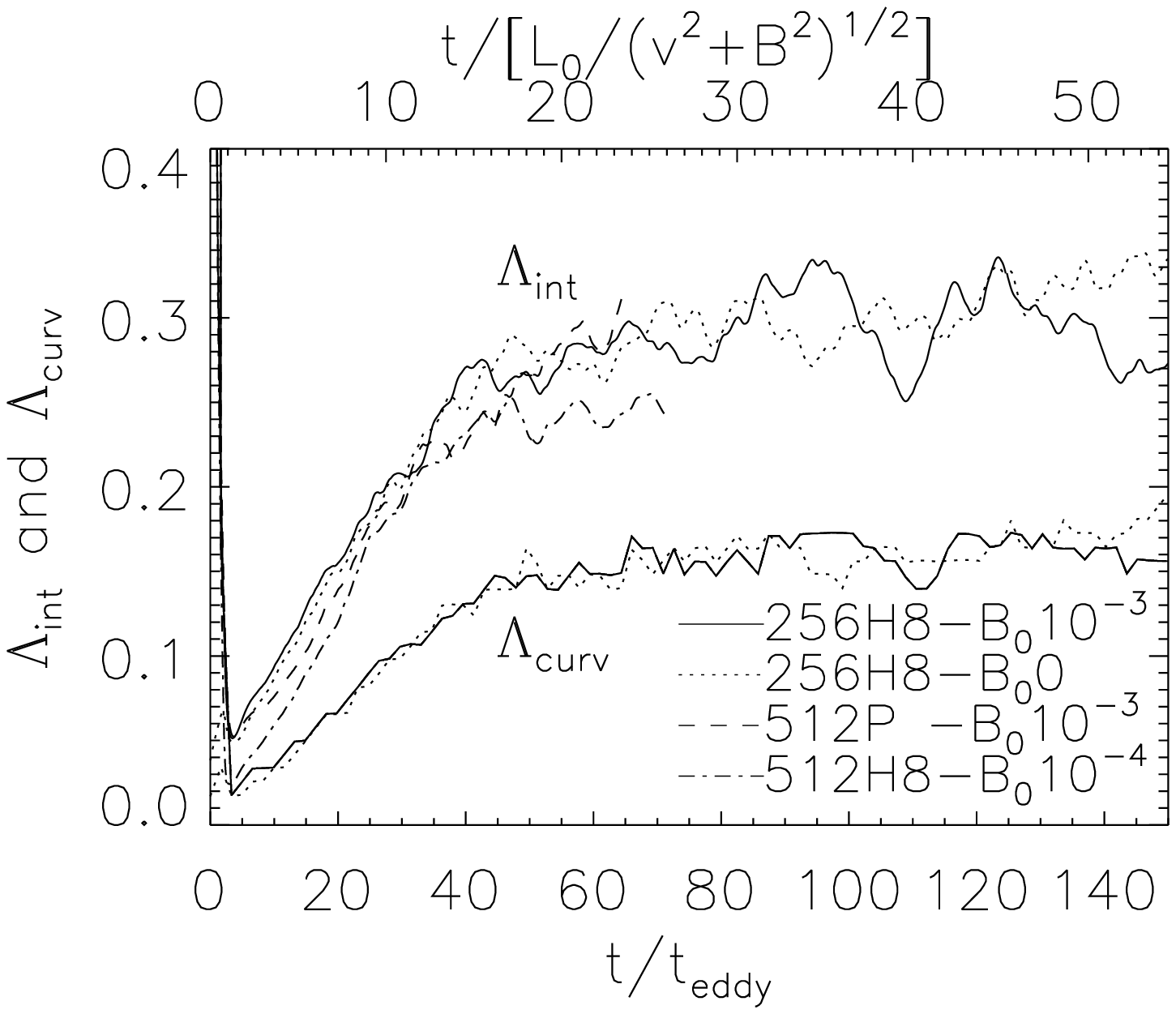}
\vskip 0cm
\caption{Time evolution of the energy equipartition scale (left panel)
and the integral and curvature scales (right panel).
$\Lambda$'s are the scales normalized with $L_0$.}
\end{figure}

\end{document}